\documentclass[preprint,12pt]{elsarticle}

\usepackage{amssymb}
\usepackage{amsmath}
\usepackage{bm}

\usepackage{tabularx}
\usepackage{booktabs}
\usepackage{multirow}
\newcommand{\frc}[2]{\raisebox{2pt}{$#1$}\big/\raisebox{-3pt}{$#2$}}
\usepackage{float}

\graphicspath{{images/bulk/}{images/pore/}}

\renewcommand{\vec}[1]{\ensuremath\bm{#1}}

\journal{Computer Physics Communications}

\begin{document}

\begin{frontmatter}



\title{Linking Theoretical and Simulation Approaches to Study Fluids in Nanoporous Media: Classical Molecular Dynamics and Density Functional Theory}


\author{Mariia Vaganova\corref{cor1}}
\ead{vaganova.ma@phystech.edu}

\author[]{Irina Nesterova}
\ead{irina.nesterova@phystech.edu}

\author[]{Yuriy Kanygin}
\ead{yuriy.kanygin@phystech.edu}

\author[]{Andrey Kazennov}
\ead{kazennov@gmail.com}

\author[]{Aleksey Khlyupin}
\ead{khlyupin@phystech.edu}

\affiliation{organization={Center for Engineering and Technology of MIPT, Moscow Institute of Physics and Technology},
addressline={Institutskiy per., 9}, city={Dolgoprudny},
postcode={141701}, 
state={Moscow Region},
country={Russian Federation}}

\cortext[cor1]{Corresponding author.}


\begin{abstract}
We propose an approach that links density functional theory (DFT) and molecular dynamics (MD) simulation to study fluid behavior in nanopores in contact with bulk (macropores). It consists of two principal steps. First, the theoretical calculation of fluid composition and density distribution in nanopore under specified thermodynamic conditions using DFT. Second, MD simulation of the confined system with obtained characteristics. Thus, we investigate an open system in a grand canonical ensemble. This method allows us to investigate both structural and dynamic properties of confined fluid at given bulk conditions and do not require computationally expensive simulation of bulk reservoir. In this work, we obtain equilibrium density profiles of pure methane, ethane and carbon dioxide and their binary mixtures in slit-like nanopores with carbon walls. Good agreement of structures obtained by theory and simulation confirms the applicability of the proposed method.
\end{abstract}

\begin{graphicalabstract}
\includegraphics[width=\linewidth]{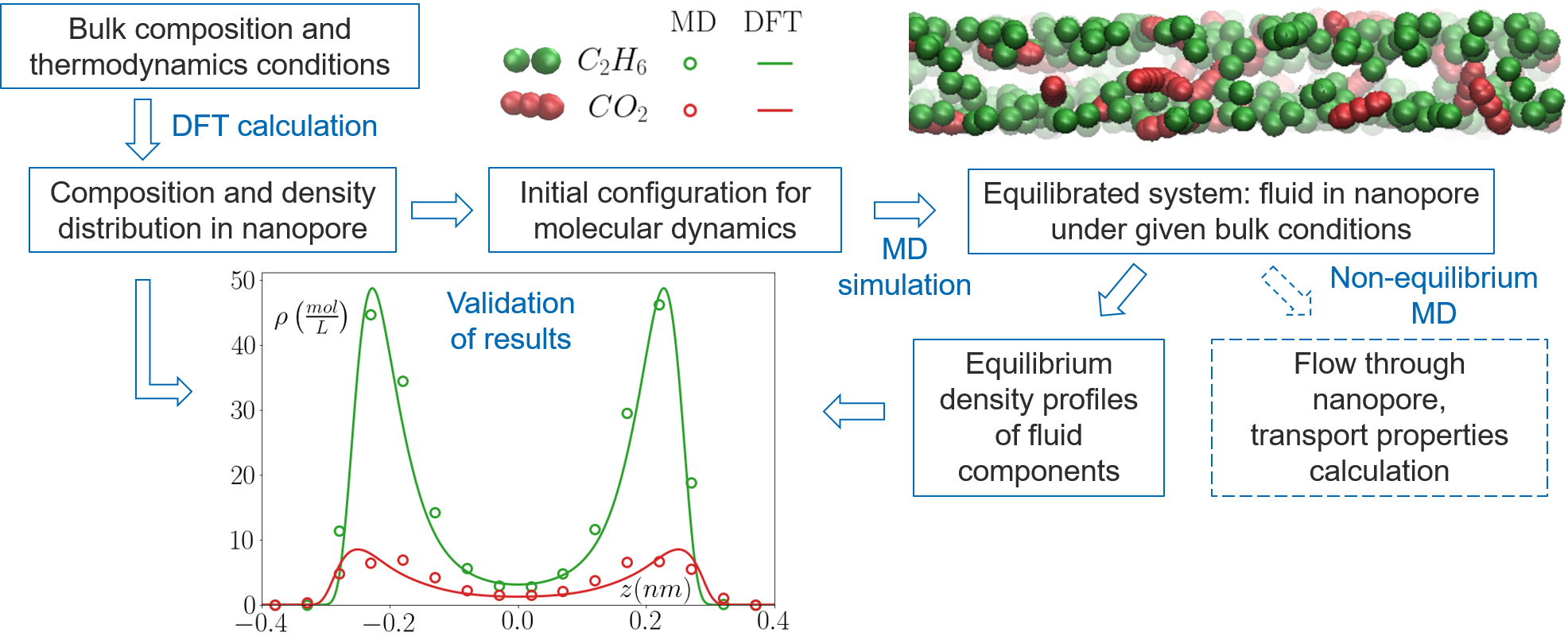}
\end{graphicalabstract}

\begin{highlights}
\item A new approach to study fluid in grand canonical ensemble is proposed
\item Our method combines molecular dynamics (MD) and density functional theory (DFT) 
\item Good agreement of simulation (MD) and theory (DFT) density profiles is obtained
\item We study competitive adsorption of hydrocarbons and carbon dioxide in nanopores
\item The proposed method can be used in further simulation to study transport phenomena
\end{highlights}


\end{frontmatter}


\section{Introduction}
\label{sec:intro}
Properties of fluid confined in nanopores that are connected with bulk (macropores) are of interest for unconventional hydrocarbon reservoirs development \cite{liu2019review,wang2019review}, geologic carbon sequestration \cite{ma2019review}, gas separation \cite{pandey2001membranes,li2009selective} and other industrial applications. It is important to develop simulation techniques for studying the adsorption, transport phenomena and phase behavior of nanoconfined fluid under given composition and thermodynamic conditions in bulk (macropores), which can be measured experimentally. 
 
Molecular dynamics (MD) simulation of fluid in nanopore is widely used to investigate its structure  \cite{santos2018molecular, tian2018understanding, sen2015molecular}, dynamic properties  \cite{wang2016fast, tang2018molecular} and phase behavior \cite{welch2015molecular, norman2016atomistic}. These studies can give insight into the fluid properties under confinement but do not provide a connection between bulk conditions and fluid inside the pore. MD simulation is a numerical integration of equations of motion for the system with a large number of interacting particles and, therefore, imposes high computational requirements.

Molecular density functional theory (DFT) is a stringent statistical physics based method, that is commonly used to investigate both micro and macroscale phenomena due to small computational requirements. DFT has been confirmed to accurately describe adsorption behaviour, capillary condensation and layering transitions and many other processes in confinement \cite{balbuena1993theoretical,neimark1998pore,ravikovitch2001density,neimark2003bridging,wu2006density,peng2008density, aslyamov2019theoretical}.

Another common technique to study adsorption in nanopores is Grand canonical Monte Carlo (GCMC) simulation. It describes the fluid in nanopore which is in equilibrium with a bulk system with an infinite number of fluid molecules and shows good agreement with experimental adsroption data \cite{kaneko1994nitrogen, heuchel1999adsorption}. However, this method is computationally expensive and can have sampling problems for low concentration components.

Both DFT and GCMC are not applicable to investigate dynamic properties. 
Thus, a different method is required to describe both thermodynamic equilibrium of nanopore with bulk and transport of fluid in nanopore.

Two approaches to this problem are found in the literature. First, MD simulation of system "bulk + pore" can be performed \cite{elola2019preferential, le2015propane}. It requires sufficiently large number of particles and is computationally expensive.
Second, combination of GCMC and MD methods is implemented in some studies (e.g. by Zhou and Wang \cite{zhou2000adsorption} to study carbon dioxide adsorption and diffusion in slit carbon pores and by Sui et al. \cite{sui2020molecular} to study flow characteristics of n-alkanes in dolomite pores).

We propose another approach that combines DFT and MD. It consists of 1) DFT calculation of the pore filling (the equilibrium density profile of each fluid component) at given bulk conditions and 2) MD simulation of obtained confined system.

This approach allows us to effectively reproduce Grand canonical ensemble and investigate both thermodynamic and transport properties of confined fluid and is more computationally efficient than MD simulation of complex "bulk + pore" system or hybrid GCMC/MD method.

In this work we compare density profiles obtained by DFT and MD to test the applicability of the proposed method. There are already some comparisons in the literature showing good agreement of adsorbed fluid density profiles obtained by DFT and MD for single-component model Lennard\,--\,Jones fluid \cite{liu2019adsorption} and binary Lennard\,--\,Jones fluid mixture \cite{sokolowski1990lennard}. In this work we considered hydrocarbons and carbon dioxide and their mixtures that are relevant for the petroleum industry (in particular, for the $CO_2$-enhanced gas recovery).

We study methane, ethane, carbon dioxide and their binary mixtures in slit-like nanopores with carbon walls. Interaction with walls is modeled by the 9-3 Lennard\,--\,Jones potential in both theory and simulation. 
We consider conditions typical for reservoirs (bulk pressure 1--10~MPa, temperatures 273--373~K), and pore widths 1, 3, 5 (nm).

The paper is organized as follows. In section \ref{sec:methods} we provide molecular models (the selection of proper force field for MD and parametrization of DFT), description of theoretical method and simulation details. In section \ref{sec:results} we present the equilibrium density profiles for single-component fluids and binary mixtures and compare DFT and MD results.

All MD simulations were performed in GROMACS package (version 2019.4) \cite{abraham2015gromacs}. Snapshots of simulated systems were produced using VMD 1.9.3 \cite{HUMP96}.

\section{Methods}
\label{sec:methods}

\subsection{Molecular Models}

Here we first provide details of MD simulation for the calculation of bulk fluid properties and the choice of force field. Then we describe DFT implementation used in this work. Approach to study confined fluid is presented in the next subsection.

\subsubsection{MD Force Fields}
\label{forcefields}

To choose force fields applicable to describe the behavior of methane, ethane and carbon dioxide under typical reservoir conditions we have compared several models, widely used in the literature. 

For carbon dioxide molecule we considered two versions of Harris and Yung model \cite{harris1995carbon}~--- fully rigid EPM2 and flexible EPM2 (with rigid bonds but flexible angle), and the fully flexible model developed by Cygan et al. \cite{cygan2012molecular}.

Rigid model with angle $180 ^{\circ}$ was constructed using virtual sites: it consists of two massive particles (arranged so that the total mass and moment of inertia of the molecule are preserved) and three virtual sites with parameters of $C$ and $O$ atoms.

For methane and ethane we have compared two force fields~--- the all-atom OPLS-AA \cite{jorgensen1996development} and the united atom TraPPE-UA \cite{martin1998transferable}.

Using this models we calculated isotherms for methane at supercritical temperature (273~K) and ethane and carbon dioxide at both sub- (273~K) and supercritical (320~K) temperatures.  Pressures up to 20~MPa were considered. Subcritical methane was not studied as its critical temperature (190.6~K) is significantly lower than the temperature range we are interested in. The results were compared with experimental data from National Institute of Standards and Technology (NIST) database \cite{nist}.

LJ parameters for interaction between unlike atoms are obtained with geometric mixing rule ($ \varepsilon_{ij} = \sqrt{\varepsilon_{ii} \varepsilon_{jj}}, \;
\sigma_{ij} = \sqrt{\sigma_{ii} \sigma_{jj}}
$)
in OPLS and EPM2 model, and with Lorentz\,--\,Berthelot mixing rule ($ \varepsilon_{ij} = \sqrt{\varepsilon_i \varepsilon_j}, \;
\sigma_{ij} = (\sigma_i + \sigma_j) / 2
$) in TraPPE and flexible $CO_2$ model. For atoms from different molecules (in mixtures) we used Lorentz\,--\,Berthelot mixing rule.

All MD simulations were performed in canonical NVT ensemble. 
For a given density value we constructed the cubic box containing 10 000 molecules and applied 3d periodic boundary conditions.
The Newton's equations were integrated by leap-frog algorithm with 1~fs time step.
Cutoff distance of 1.4~nm was set for LJ potential and long-range dispersion corrections for energy and pressure \cite{allen2017computer} were applied.
Temperature was coupled using V-rescale thermostat \cite{bussi2007canonical} with time constant of 0.1~ps.
For systems with charges long-range electrostatic interactions were calculated using Particle Mesh Ewald (PME) method with 1.4~nm cutoff.
Linear Constraint Solver (LINCS) algorithm \cite{hess1997lincs} was used to constraint all bonds in TraPPE and EPM2 and bonds which include hydrogen in OPLS.

\begin{figure}[h!]
\centering
    \includegraphics[width=0.5\linewidth]{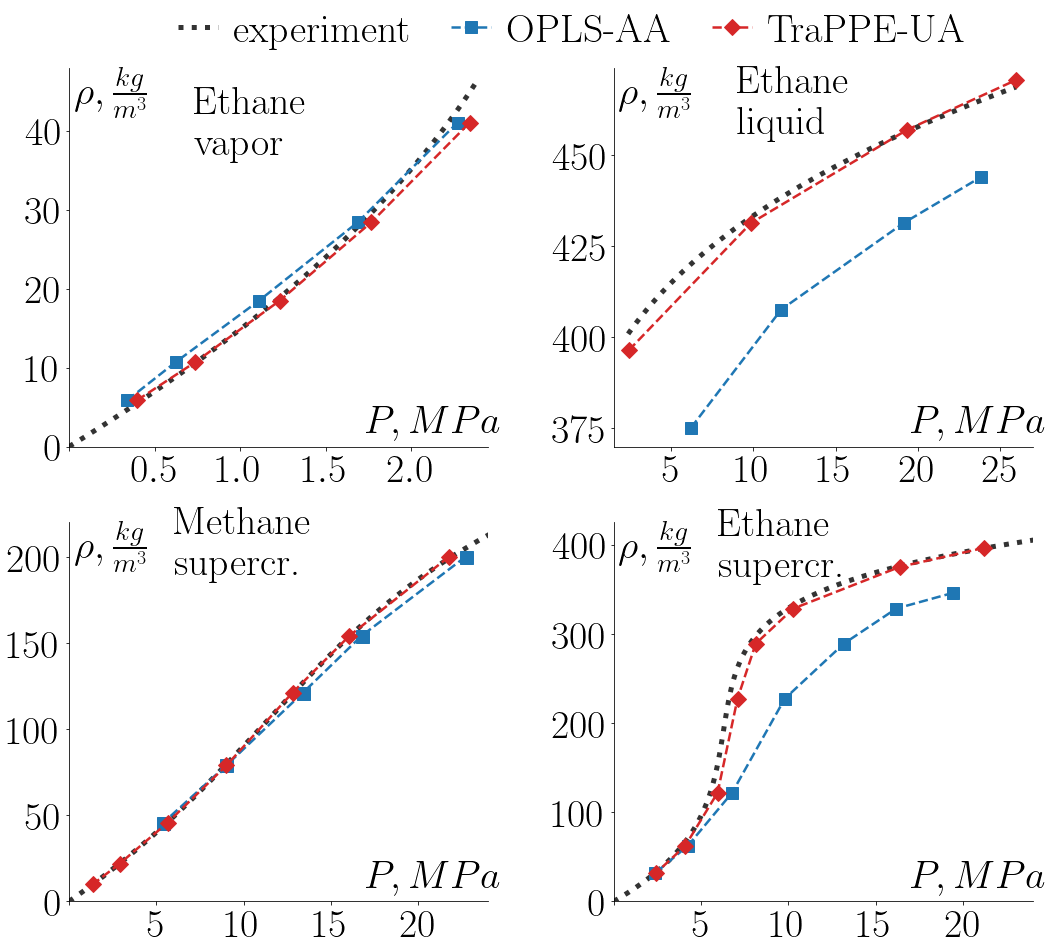}
    \caption{Isotherms for methane and ethane in bulk. Comparison of MD results obtained using different force fields with experimental data from NIST}
    \label{fig:meth_eth_ff}
\end{figure}

\begin{figure}[h!]
\centering
    \includegraphics[width=0.5\linewidth]{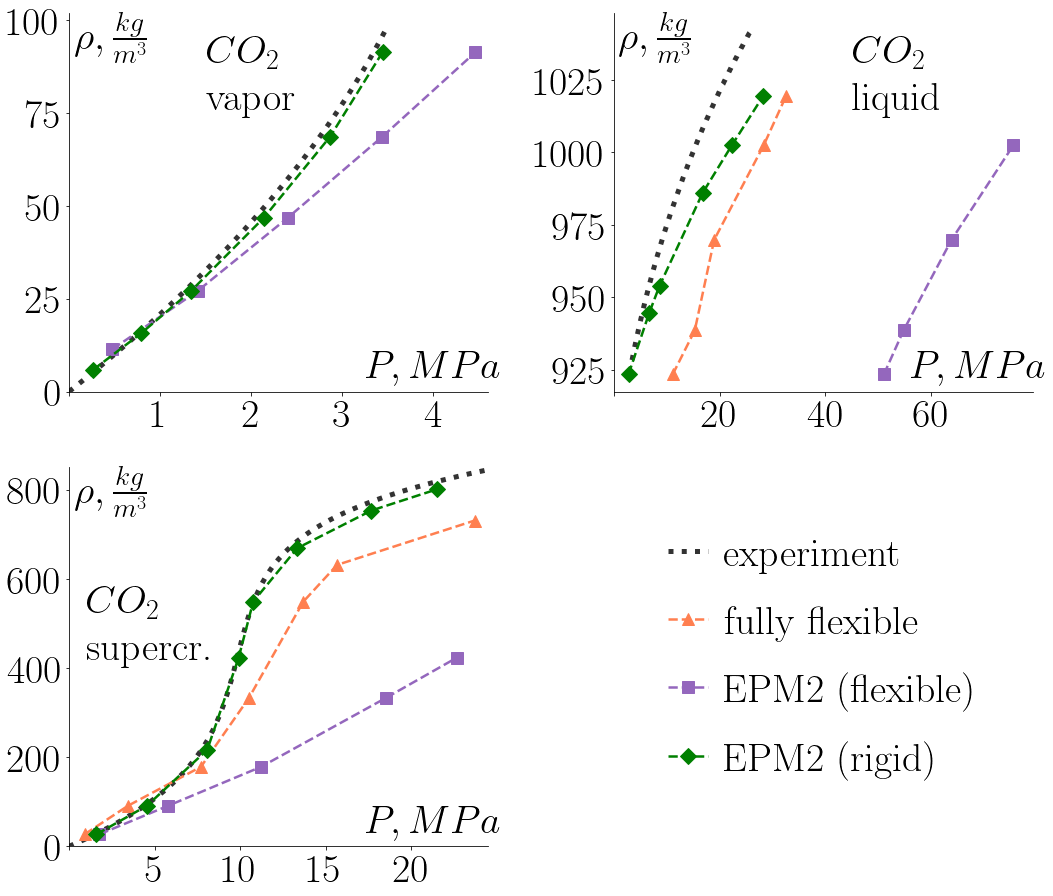}
    \caption{Isotherms for carbon dioxide in bulk. Comparison of MD results obtained using different force fields with experimental data from NIST}
    \label{fig:co2_ff}
\end{figure}

For each system 2~ns run was performed. If the system did not reach equilibrium by this time (some systems with flexible molecules), the run was continued. In equilibrium average pressure was obtained from the last 500~ps. The resulting density-pressure dependencies are shown in figures \ref{fig:meth_eth_ff}\,--\,\ref{fig:co2_ff}.

Some flexible models turned out to be unsuitable for simulations of low-density systems. Namely, in simulations of methane at $ \lesssim 20 \; kg/m3$ with OPLS and of carbon dioxide at 273~K at $ \lesssim 100 \; kg/m3$ with fully flexible model the system did not reach equilibrium and droplets formed in the gas.

Rigid EPM2 model for carbon dioxide and TraPPE-UA force field for methane and ethane showed better agreement with experiment (as can be seen from the figures) and were chosen for further research.

\subsubsection{DFT Bulk}

Classical density functional theory is widely used statistical-mechanical method that bridges macroscopic systems properties with molecular phenomena. DFT provides bulk equation of state (EoS) that is determined by parameters of intermolecular interactions. In this work we account short ranged "hard-sphere" interactions and long ranged "attraction" interactions using fundamental measure theory (FMT) and mean-field approximation (MFA), respectively. To obtain bulk EoS, we should consider canonical ensemble, where Helmholtz Free energy in terms of DFT is formulated as:

\begin{gather}
    F\left[\rho\right] =  F^{id}\left[\rho\right]+F^{hs}\left[\rho\right]+ F^{att}\left[\rho\right],\label{eq:f_sum_mix}\\
    F^{id}\left[\rho\right] = k_B T \sum_{i}\int d\vec{r}\,\rho_i\left(\vec{r}\right)\left(\ln{{(\Lambda_i}^3 \rho_i\left(\vec{r}\right))}-1\right),\label{eq:f_id_mix}\\
    F^{hs}\left[\rho\right] = k_B T\int d\vec{r}\,\Phi\left[n_\alpha\left(\vec{r}\right)\right],\label{eq:f_hs_mix}\\
    F^{att}\left[\rho\right] = k_B T \sum_{i,j}\int \int d\vec{r}\rho_i\left(\vec{r}\right)d\vec{r}^\prime\rho_j\left(\vec{r}^\prime\right)U_{ij}^{att}(\vert\vec{r}-\vec{r}^\prime\vert) \label{eq:f_att_mix}.
\end{gather}
where $i$ is the component index, $\Lambda_i = \frc{h}{\sqrt{2\pi m_iT}}$ is the thermal de Broglie wavelength, $\Phi\left[n_\alpha\left(\vec{r}\right)\right]$~--- Rosenfeld functional, that takes the form:
\begin{equation}\label{eq:rosienfield}
    \Phi =  -n_0 \ln{\left(1-n_3\right)} + \frac{n_1 n_2- \vec{n_1}\cdot\vec{n_2}}{1-n_3} + \frac{n_2^3-3n_2 \vec{n_2}\cdot\vec{n_2}}{24\pi\left(1-n_3\right)^2}.   
\end{equation}

Here functions $n_\alpha, \bm{n}_\beta$ are the scalar and vector weighted densities, respectively ($\alpha = 0,1,2,3;\,\beta = 1,2$), defined as:
\begin{gather}\label{eq:weighted_dens}
    n_\alpha \left(\vec{r}\right)=\sum_{i} \int d^3r^\prime \rho_i\left(\vec{r}^\prime\right)\omega_\alpha^i\left(\vec{r}-\vec{r}^\prime\right),\\
    \bm{n}_\beta \left(\vec{r}\right)=\sum_{i} \int d^3r^\prime \rho_i\left(\vec{r}^\prime\right),\bm{\omega}_\beta^i\left(\vec{r}-\vec{r}^\prime\right).
\end{gather}

Weight functions $\omega^i_\alpha, \bm{\omega}^i_\beta$ are given by : $\omega^i_3\left(\vec{r}\right)=\theta\left(R_i-r\right)$, $\omega^i_2\left(\vec{r}\right)=\delta\left(R_i-r\right)$,  $\omega^i_1= \frc{\omega^i_2}{4\pi R_i}$, $\omega^i_0 = \frc{\omega^i_2}{4\pi R_i^2}$,
${\vec{\omega}}^i_2\left(\vec{r}\right)=\frac{\vec{r}}{r}\delta\left(R_i-r\right)$,
${\vec{\omega}}^i_1= \frc{{\vec{\omega}}^i_2}{4\pi R_i}$. Here $\delta$ and $\theta$ are Dirac-delta function and Heaviside step function, respectively, and $R_i$ is the radius of the spherical model of the molecule.

The potential of attraction interactions $U^{att}$ is split according to the WCA scheme \cite{weeks1971role}:

\begin{equation}
    U_{ij}^{att}\left(r\right)=\ \left\{
    \begin{matrix}
        -\epsilon_{ij}, & r<\lambda_{ij}\\
        U_{ij}^{LJ}, & \lambda_{ij}<r<r_{cut}\\
        0, & r>r_{cut}\\
    \end{matrix}\right.
\end{equation}

\begin{equation}
    U_{ij}^{LJ}=4\epsilon_{ij}\left(\left(\frac{\sigma_{ij}}{r}\right)^{12}- \left(\frac{\sigma_{ij}}{r}\right)^6\right)
\end{equation}

Here $r = \vert\vec{r}-\vec{r}^\prime\vert$ is the distance between two fluid molecules, $\epsilon_{ij}$ and $\sigma_{ij}$ are Lennard\,--\,Jones intermolecular interaction parameters, $\lambda_{ij} = 2^{1/6} \sigma_{ij}$ is the location of Lenard\,--\,Jones potential minimum and $r_{cut}$ is the cutoff distance. We consider $r_{cut} = \infty$ and $\sigma_{ii}=2R_i$.

In the limit of $H \to \infty$, particle distribution function reduces to constant function $\rho_i(\vec{r}) \to \rho_i$, so Helmholtz Free energy functional (eq. \ref{eq:f_sum_mix} -- \ref{eq:f_att_mix}) turns into function of $\rho_i$. Then, fluid pressure-density relation  and chemical potential for each component can be derived.

\begin{gather}
    p = p^{id} +p^{hs} + p^{att}\label{eq:p_2k}\\
    p^{id} = \sum_{i}\rho_i k_B T \label{eq:p_2k_id}\\
    p^{hs} = \sum_{i}\rho_i k_B T \left(\frac{1 + 2\eta+3\eta^2}{(1-\eta)^2}-1\right)\label{eq:p_2k_hs}\\
    p^{att} = 0.5 \sum_{i,j}\rho_i \rho_j k_B T\int d\vec{r} U_{ij}^{att}(\vec{r})\label{eq:p_2k_att}
\end{gather}
Here $\eta = \sum_{i}\eta_i$ is the fluid packing fraction with $\eta_i = \frac{4}{3} \pi R^3 \rho_i$~--- the packing fraction of component $i$. 

\begin{gather}
    \mu_i = \mu_i^{id} +\mu_i^{hs} + \mu_i^{att}\label{eq:mu_2k}\\
    \mu_i^{id} = k_B T \ln{\Lambda_i^3 \rho_i}\label{eq:mu_2k_id}\\
    \mu_i^{hs} = k_B T \bigg( \frac{\partial \Phi}{\partial n_3} \cdot \frac{4}{3}\pi R_i^3 + \frac{\partial \Phi}{\partial n_2} \cdot \pi R_i^2 +
    \frac{\partial \Phi}{\partial n_1} \cdot R_i+ \frac{\partial \Phi}{\partial n_0}\bigg) \label{eq:mu_2k_hs}\\
    \mu_i^{att} = k_B T \rho_i\int d\vec{r} U_{ii}^{att}(\vec{r}) + \sum_{j}k_B T \rho_j\int d\vec{r} U_{ij}^{att}(\vec{r}) \label{eq:mu_2k_att}
\end{gather}

Equations \ref{eq:p_2k} -- \ref{eq:p_2k_att} and \ref{eq:mu_2k} -- \ref{eq:mu_2k_att} contains intermolecular interaction parameters $\sigma_{ij}$ and  $\epsilon_{ij}$, that relates with fluid-fluid parameters of pure components $\sigma_{ii}$ and $\epsilon_{ii}$. Parameters searching procedure is based on the results of \cite{nesterova2021adaptive} (see \ref{param_app}).

Obtained Lennard\,--\,Jones parameters for methane, ethane and carbon dioxide are given in table \ref{tab:param-pure-dft}. And table \ref{tab:param_mix_dft} provides the intermolecular interaction parameters $\epsilon_{ij}$ and $\sigma_{ij}$ used for EoS of binary mixtures. Resulting DFT EoS for both single-component fluids and binary mixtures are presented in section \ref{sec:results} along with the corresponding results of MD simulations in bulk.

\begin{table}[h!]
\centering
\caption{Parameters of Lenard\,--\,Jones potential for methane, ethane and carbon dioxide at different temperatures}
\label{tab:param-pure-dft}
\begin{tabular}{c|c|c|c}

\multirow{2}{*}{Fluid}&
\multirow{2}{*}{$T, K$} &
\multirow{2}{*}{$\frc{\epsilon_{ii}}{k_B},$ K} &
\multirow{2}{*}{$\sigma_{ii},$ \AA} \\
& & &
\\
\hline
$CH_4$ & 273 & 3.530 & 138.39 \\ 
\hline
\multirow{2}{*}{$C_2H_6$}
& 273 & 4.240 & 234.64    \\ 
& 320 & 4.097 & 221.15   \\ 
\hline
\multirow{2}{*}{$CO_2$}
& 273 & 3.628 & 233.33   \\ 
& 320 & 3.511 & 219.15   \\ 
\end{tabular}
\end{table}

\begin{table}[h!]
\centering
\caption{Intermolecular parameters for binary mixtures}
\label{tab:param_mix_dft}
\begin{tabular}{c|c|c|c}

\multirow{2}{*}{Mixture}&
\multirow{2}{*}{$T, K$} &
\multirow{2}{*}{$\frc{\epsilon_{ij}}{k_B},$ K} &
\multirow{2}{*}{$\sigma_{ij},$ \AA} \\
& & &
\\
\hline
$CH_4 + C_2H_6$  & 293 & 3.900     & 174.42 \\ 
\hline
$CH_4 + CO_2$  & 373 & 3.541
& 162.21    \\ 
\hline
$C_2H_6 + CO_2$  & 320 & 3.804  
& 197.41   \\ 
\end{tabular}
\end{table}

\subsection{Confined Fluid}

The approach linking DFT and MD to study fluid in nanoporous media is presented in the section below. 
The proposed method is aimed at describing both thermodynamic and transport behavior of fluid confined within nanopore which is in contact with bulk.

The use of MD simulation is crucial for further study of transport behavior (that is impossible by the means of DFT). And the theoretical calculation providing equilibrium with bulk makes our method computationally efficient.

Here we provide the algorithm for fluid mixture. The algorithm for single-component fluid is the same, except for calculating the composition. Further, we provide the details of theory and simulation used in our approach in sections \ref{DFT_pore} and \ref{MD pore}, respectively.

\textit{Input:} fluid composition in bulk, temperature, bulk pressure, pore width.

\begin{enumerate}
    \item Obtain component composition inside the pore and equilibrium density profile of each fluid component  $\rho_i \left( z \right)$ under given conditions in bulk and given bulk concentrations using DFT.
    \item Generate initial configuration for MD simulation with the same number of molecules of each component per unit area $N_i/S = \int_{-H/2}^{H/2} \rho_i \left( z \right) dz$.
    \item Perform NVT equilibration using MD and obtain density profiles of fluid components (averaged by the pore area).
    \item Validate the results by comparing density profiles obtained by MD and DFT.
\end{enumerate}

\textit{Output:} density profiles of each component inside the pore, system for further MD simulations.

\begin{enumerate}
    \setcounter{enumi}{4}
    \item (In perspective) Perform further MD simulations to study transport properties of confined fluid.
\end{enumerate}

The important feature of this algorithm is that though MD simulation only of fluid inside the pore is performed, at the output a density profile is obtained, which is realized at the equilibrium of the pore with bulk. It is provided by the use of input data from DFT calculation and can be confirmed by comparing theoretical and simulation results.

In both MD and DFT calculations slit pore is modeled by two carbon walls at $z = -H/2$ and $z = H/2$.
Interaction between fluid particles and each wall is represented by 9-3 Lennard\,--\,Jones potential
$$
V_{wall}(z) = 2 \pi \rho_s \epsilon_{sf} \sigma_{sf}^3
\left[ \frac{2}{45}\left( \frac{\sigma_{sf}}{z}\right)^9 -\frac{1}{3}\left( \frac{\sigma_{sf}}{z}\right)^3
\right]
$$
with carbon parameters $\sigma_{ss} = 0.34 \; nm$, $\epsilon_{ss}/k_B = 28 \; K$, $\rho_s = 114 \; nm^{-3}$ \cite{steele1973physical}. 
Parameters for solid-fluid interaction are obtained by Lorentz\,--\,Berthelot mixing rule (for each atom in MD and for each molecule in DFT). The total external potential is
$$
V_{ext}(z) = V_{wall}(z-H/2) + V_{wall}(z+H/2).
$$

\subsubsection{DFT Pore}
\label{DFT_pore}
We consider pore finite along z axis, and symetric along x and y axes.
According to DFT, intrinsic Helmholtz free energy $F$ can be expressed as the functional of density. Free energy of Grand Canonical ensemble (system pore + bulk) is $\Omega$ potential:

\begin{equation}\label{eq:Omega}
    \Omega\left[\rho\right]=F\left[\rho\right]+\int d\vec{r} \sum_{i} \rho_i\left(\vec{r}\right)\left(V_i^{ext}\left(\vec{r}\right)-\mu_i\right),
\end{equation}
where $i$ is the component index, $V_i^{ext}$ is the external potential, and $\mu_i$ is the chemical potential. Thermodynamic potential is to be minimized to find equilibrium state of system. Implementing minimization procedure, one can obtain $\rho_i\left(\vec{r}\right)$ :

\begin{equation}\label{eq:density_dft}
    {\rho_i}\left(\vec{r}\right) = \rho_i^{bulk}\exp{\left\lbrace -\frac{1}{k_B T}\left( \dfrac{\delta F\left[\rho\right]}{\delta \rho_i\left(\vec{r}\right)} + V_i^{ext}\left(\vec{r}\right) - \mu_i^{ex}\right)\right\rbrace},
\end{equation}
where $\rho_i^{bulk}$ is the bulk fluid density of component $i$, $k_B$ is Boltzmann constant, $T$ is system temperature, $F \left[\rho\right]$ is given by eq. \ref{eq:f_sum_mix}--\ref{eq:f_att_mix}, $\mu_i^{ex} = \mu_i - \mu_i^{id}$ is the excess chemical potential, which can be determined from \ref{eq:mu_2k}--\ref{eq:mu_2k_att}.

Equilibrium density searching procedure is implementing with Picard iteration method. After $\rho_i \left( z \right)$ is obtained we integrate it along pore width and give the average density as an income to MD simulation.

\subsubsection{MD Pore}
\label{MD pore}

For MD simulation of fluid in the pore we create the 3d system finite along $z$ axis and apply periodic boundary conditions along $x$ and $y$ axes. Height of the simulation box equals to the pore size (H = 1~nm, 3~nm or 5~nm) and width $L_x=L_y$ is in range of 25--60~nm.

First, the average number of molecules (of each fluid component) per unit area inside the pore is obtained from DFT calculations: $N_i/S = \int_{-H/2}^{H/2} \rho_i \left( z \right) dz$. Then we select the pore area value such that the total number of molecules in the system is between 10000 and 80000. To avoid finite size effect we have compared results obtained with different number of particles (in some cases with droplet formation 10 000 molecules were not enough).

Initial molecule positions are generated randomly and some atoms may be beyond the wall. For this reason parameter "wall-r-linpot" (the distance from the wall below that potential is continued linearly) is set positive ($\sim 0.1 \; nm$) during the potential energy minimization. After energy minimization this parameter is set to -1, that means that atoms can not be beyond a wall (in this case a fatal error is generated).

Long range electrostatic interactions are calculated using a modified 3d Particle Mesh Ewald method with corrections to force and potential in the $z$ dimension \cite{yeh1999ewald} which performs a pseudo-2D summation.

Other parameters for MD simulations are the same as described in section \ref{forcefields}.

We performed 10~ns NVT-equilibration run for each system and used data from the last 2~ns to calculate equilibrium density profile. It was obtained by dividing $z$-axis into slices 0.05~nm thick and calculating the average number of molecules in each slice (by positions of molecules' center of mass). Averaging was performed both over time and over the pore area.

\section{Results}
\label{sec:results}

\subsection{Single-Component Fluid}
\label{pure_pore}

\begin{figure}[h!]
\centering
\includegraphics[width=0.5\linewidth]{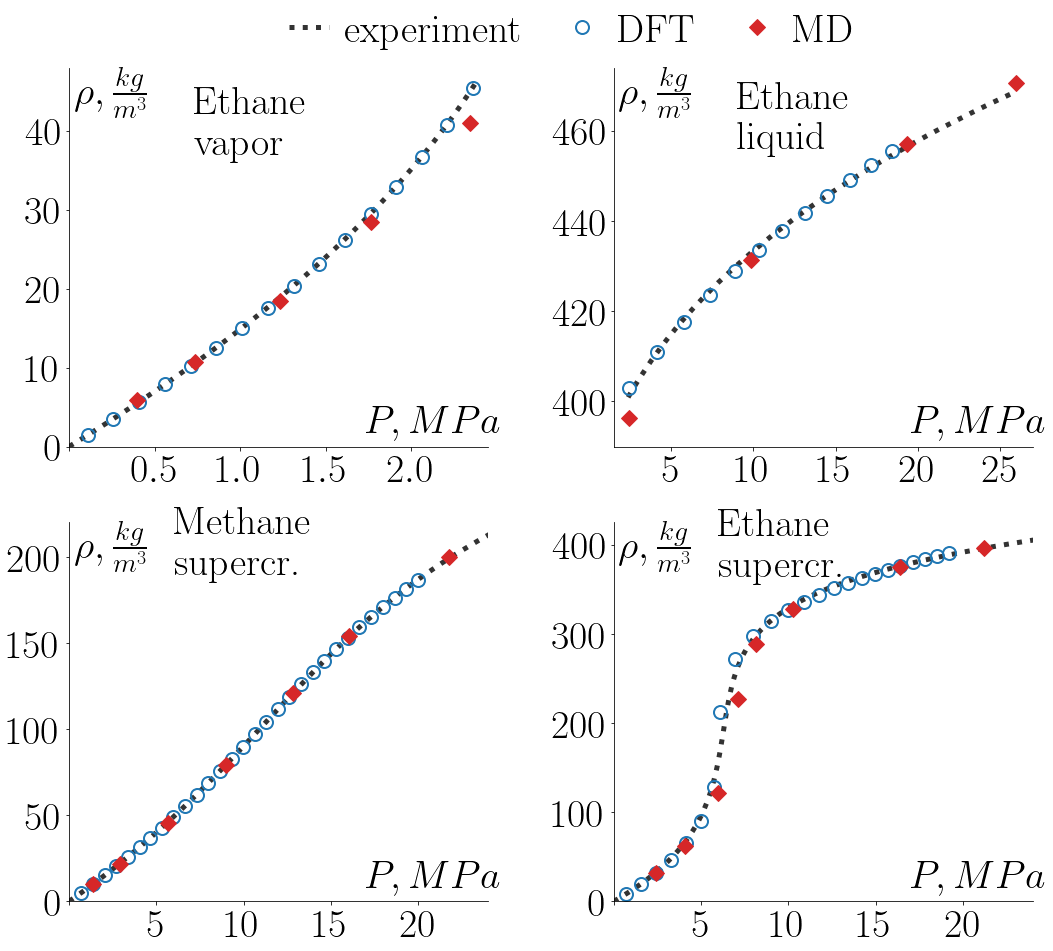}
\caption{Isotherms for methane and ethane in bulk. Comparison of MD and DFT results with experimental data from NIST}
\label{fig:meth_eth_pure}
\end{figure}

\begin{figure}[h!]
\centering
\includegraphics[width=0.5\linewidth]{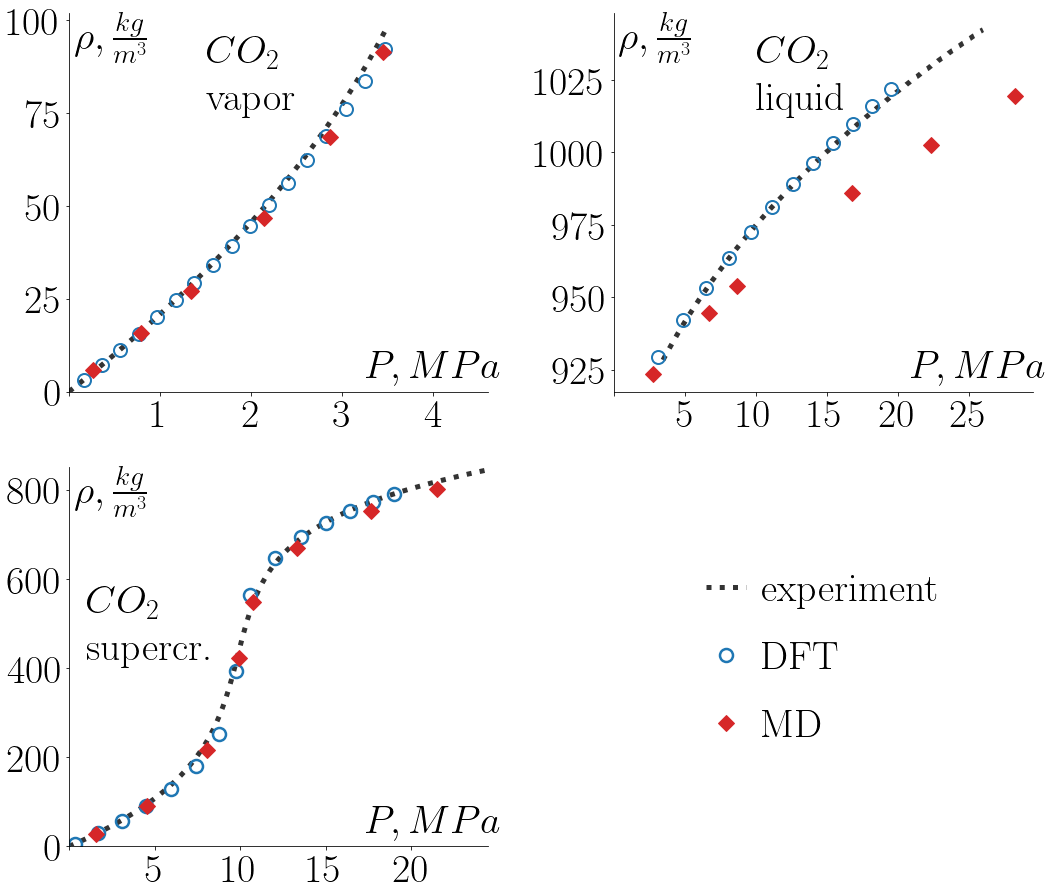}
\caption{Isotherms for carbon dioxide in bulk. Comparison of MD and DFT results with experimental data from NIST}
\label{fig:co2_pure}
\end{figure}

Before applying our method to confined fluid we validated MD and DFT models on bulk PVT fluid properties.
Figures \ref{fig:meth_eth_pure}, \ref{fig:co2_pure} shows isotherms for methane at 273~K, and both subcritical (273~K) and supercritical (320~K) ethane and carbon dioxide at pressures up to 20~MPa. As we can see dependencies obtained by MD and DFT are in good agreement with each other and with experimental data from NIST \cite{nist}. The density deviation increases at high pressures and is less than 3 \% in the whole range studied.

Then we applied our approach to a single component fluid in slit nanopores of different widths and compared equilibrium density profile obtained by MD and DFT. We studied methane at 273~K and ethane and carbon dioxide at both sub- and supercritical temperatures (273~K and 320~K, respectively). In each case we fixed bulk pressure (chosen so that fluid in bulk was in gaseous state) and investigated the filling of pores with widths of 1, 3 and 5 (nm).

\begin{figure}[h!]
    \centering
    \includegraphics[width=0.5\linewidth]{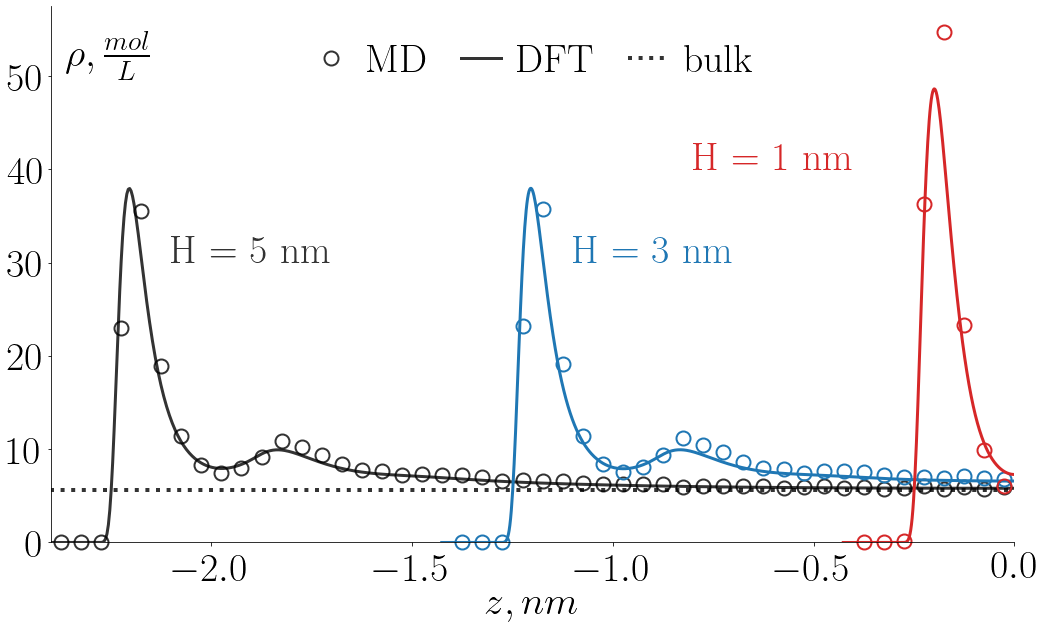}
    \caption{Methane density profiles, T = 273 K, bulk pressure 10 MPa}
    \label{fig:meth_pore}
\end{figure}

\begin{figure}[h!]
    \centering
    \includegraphics[width=0.5\linewidth]{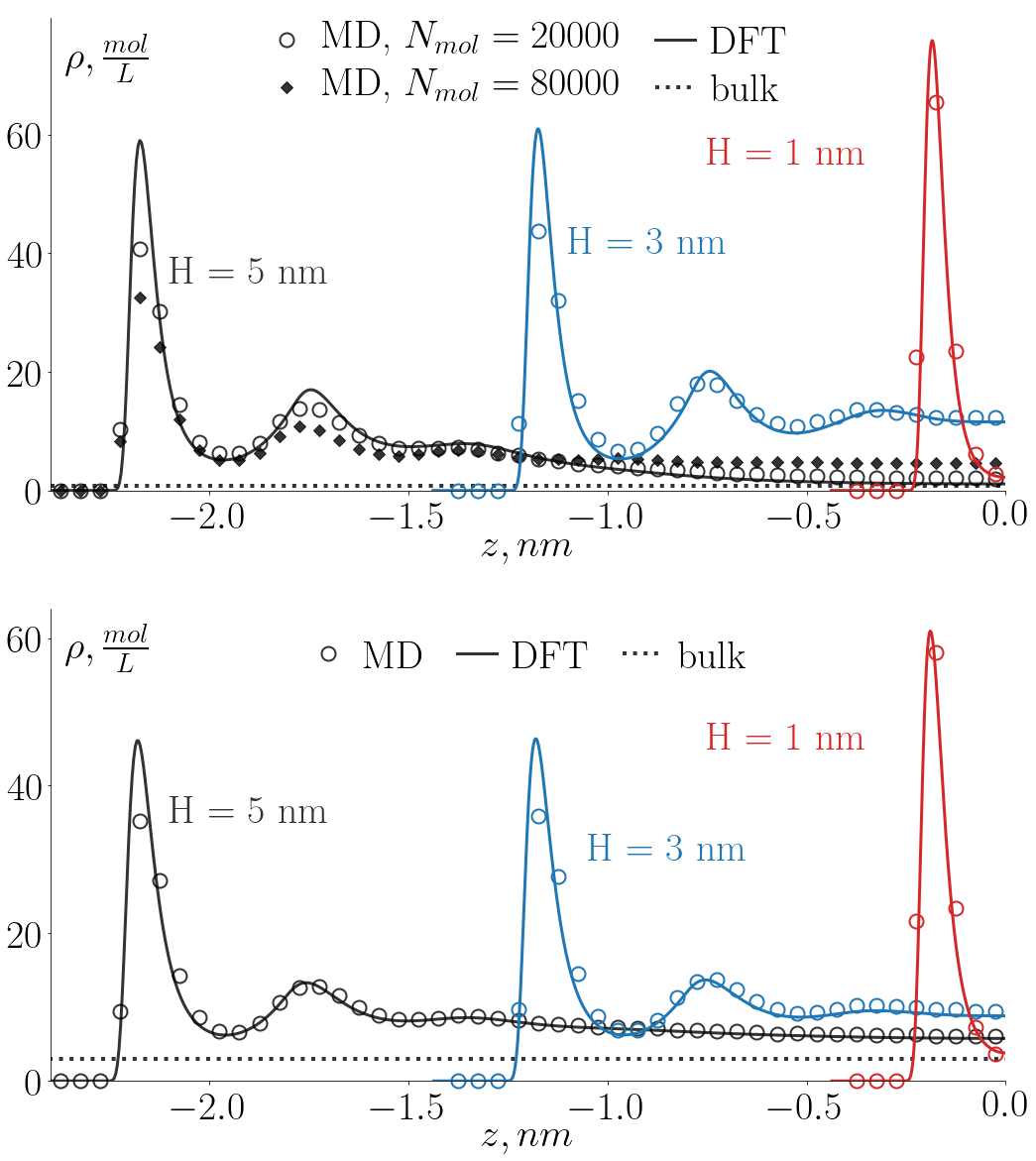}
    \caption{Ethane density profiles, top: T = 273 K, bulk pressure 1.5 MPa; bottom: T = 320 K, bulk pressure 5 MPa}
    \label{fig:eth_pore}
\end{figure}

\begin{figure}[h!]
    \centering
    \includegraphics[width=0.5\linewidth]{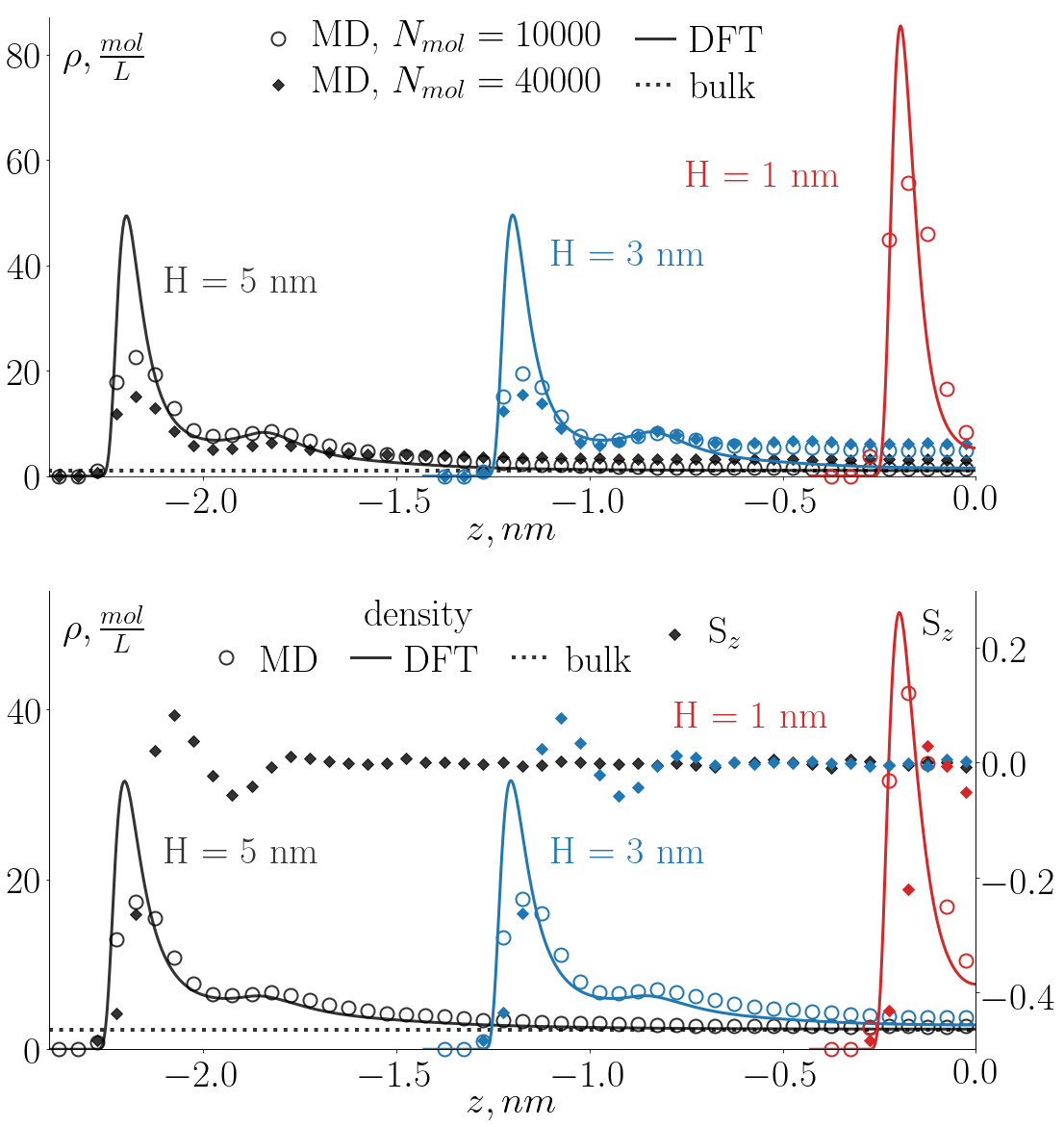}
    \caption{Carbon dioxide density profiles and order parameter dependence on z, top: T = 273 K, bulk pressure 2 MPa; bottom: T = 320 K, bulk pressure 5 MPa}
    \label{fig:co2_pore}
\end{figure}

Comparison of density profiles obtained by MD and DFT is shown in figures \ref{fig:meth_pore}, \ref{fig:eth_pore} and \ref{fig:co2_pore} for methane, ethane and carbon dioxide, respectively. Z-distance is calculated from the center of the pore and only the half of the pore (from $z = -H/2$ to $z = 0$) is presented in the figures as all profiles are symmetrical.

In all cases the structure of profiles agrees well~--- we observe the formation of the equal number of adsorption layers and (in the pores wider than 1 nm) the bulk-like region with constant density in the center of the pore. The positions and thickness of the layers predicted by MD and DFT are also very close.

For methane in 3 and 5~nm pores (fig.~\ref{fig:meth_pore}) and for ethane in 1~nm pores (fig.~\ref{fig:eth_pore}) MD and DFT profiles show very good qualitative and quantitative agreement.

In the other cases discrepancy is observed for the peak value of density in the first adsorption layer. MD peak is higher than DFT for methane in 1~nm pore (fig. \ref{fig:meth_pore}). DFT peak is higher than MD for ethane in 3 and 5~nm pores (fig. \ref{fig:eth_pore}) and in all cases for carbon dioxide (fig. \ref{fig:co2_pore}).

The difference in density profiles can be explained by the different molecular models used by DFT and MD. In DFT a molecule of a real fluid is represented as hard sphere with the effective long-range attraction potential, which most accurately describes the behavior of this fluid in bulk. Because of this difference, DFT does not reproduce so precisely a real fluid behavior under confinement. Inaccuracies due to mean-field treatment of attractive molecular forces and contribution of hard-spheres repulsion have already been observed by other studies \cite{sokolowski1990lennard}.

In both MD and DFT profiles for ethane in 3~nm pore at 273~K and in 3 and 5~nm pores at 320~K (fig. \ref{fig:eth_pore}) and for carbon dioxide in 1~nm pores (fig. \ref{fig:co2_pore}) density in the center of the pore is higher than bulk density, which means that capillary condensation has occurred. In these cases fluid is in the gaseous state in bulk, but inside the pore it becomes liquid.

Under certain conditions (for ethane in 5~nm pore at 273~K and for carbon dioxide in 3 and 5~nm pores at 273~K) we observed the process of capillary condensation in MD simulation. As shown in figures \ref{fig:eth_pore}, \ref{fig:co2_pore} with increasing number of molecules in the system density in the center of the pore increases and becomes larger than bulk density. If the number of molecules is large enough the droplet is formed inside the pore. DFT in these cases also did not provide an unambiguous solution~--- it converged to gas or liquid density profiles depending on the initial approximation. The figures above show only gas profiles (with an initial density equal to the bulk density) and we don't discuss these cases here as this is not the subject of current work. However, it shows the possibility to apply our method to phase transitions in nanopores which will be interesting to investigate more thoroughly in the future. 

Table \ref{tab:rmse_pure} presents quantitative comparison between theoretical and simulation results (for the cases without droplets formation). Discrepancy between density profiles is calculated as normalized root-mean-square deviation:
$$NRMSD = \frac{
\sqrt{
\frac{1}{N_{sl}}
\sum_{i=1}^{N_{sl}} \left(\rho^{DFT}_i - \rho^{MD}_i\right)^2}
}
{\max_i\left\{\rho_i^{DFT} + \rho_i^{MD}\right\}/2},$$
where $N_{sl} = H / \Delta z$ is the number of $z$-axis slices by which the density profile was calculated and $\rho_i$ is the density value at $z = i\Delta z$ point.

\begin{table}[h!]
\centering
    \caption{Normalized RMSD between MD and DFT density profiles for single-component fluids}
    \label{tab:rmse_pure}
\begin{tabular}{c|c|c|c}
    Fluid & T, K & 
    H, nm & NRMSD, \% \\
    \hline
    \multirow{3}{*}{$CH_4$} & \multirow{3}{*}{273} 
    & 1 & 8.8 \\
    & & 3 & 4.0 \\
    & & 5 & 2.7 \\
    \hline
    \multirow{5}{*}{$C_2H_6$} & \multirow{2}{*}{273} 
    & 1 & 9.0 \\
    & & 3 & 7.4 \\
    \cline{2-4}
    & \multirow{3}{*}{320} 
    & 1 & 6.4 \\
    & & 3 & 5.4 \\
    & & 5 & 4.5 \\
    \hline
    \multirow{4}{*}{$CO_2$} & 273 & 1 & 13.1 \\
    \cline{2-4}
    & \multirow{3}{*}{320} 
    & 1 & 9.6 \\
    & & 3 & 10.5 \\
    & & 5 & 7.6 \\
\end{tabular}
\end{table}

As we can see both from visual comparison and from the RMSD values the discrepancy between DFT and MD results is most noticeable for carbon dioxide. It can be so because the carbon dioxide molecule is linear and its distribution near the wall surface depends on orientation which is not taken into account in DFT. 

To prove this assumption we investigated molecules' equilibrium orientation in the pore for carbon dioxide at 320~K and bulk pressure 5~MPa. We divided $z$-axis into slices and calculated the order parameter
$S_z = \frac{3}{2}\langle cos^2 \theta_z \rangle - \frac{1}{2}$, where $\theta_z$ is the angle between $z$-axis and the vector between two oxygen atoms in $CO_2$ molecule and $\langle * \rangle$ means averaging over ensemble and time in each slice. The resulting orientation dependence on $z$ along with MD and DFT density profiles is shown in fig. \ref{fig:co2_pore}.

\begin{figure}[H]
    \centering
    \includegraphics[width=0.5\linewidth]{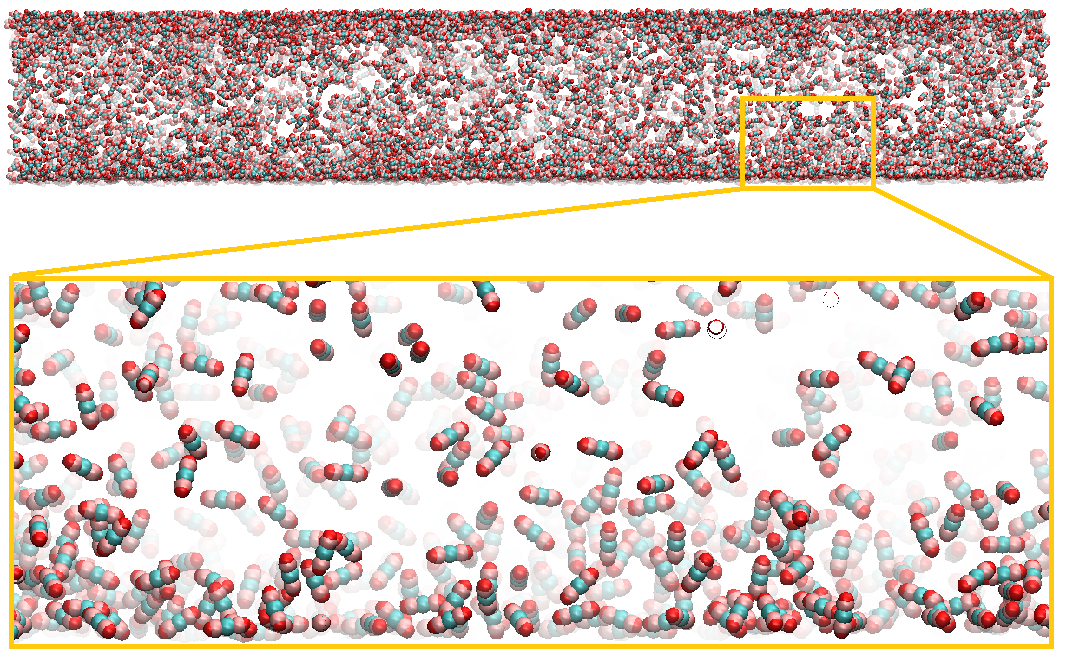}
    \caption{Carbon dioxide in 5 nm pore, T = 320 K. Snapshot from MD simulation}
    \label{fig:co2_320_5_snapshot}
\end{figure}

As we can see for all pore widths the order parameter equals 0 in the center of the pore and becomes negative in the adsorption layer. It means that molecules orient isotropically in the center of the pore but near the wall mostly orient parallel to the surface, which can be also seen in the snapshot from MD simulation for 5~nm pore (fig. \ref{fig:co2_320_5_snapshot}). Such behavior can be caused by the polarity of $CO_2$.
In DFT the representation of molecules is spherically symmetric with a radius smaller than the linear length of the molecule, so each parallel-oriented molecule occupies a larger surface area than its DFT representation and DFT predicts a larger number of molecules in the adsorption layer than MD. And since the total number of molecules is set equal, DFT predicts fewer molecules in the center of the pore.

\subsection{Binary Mixture}

\begin{figure}[h!]
\centering
\includegraphics[width=0.5\linewidth]{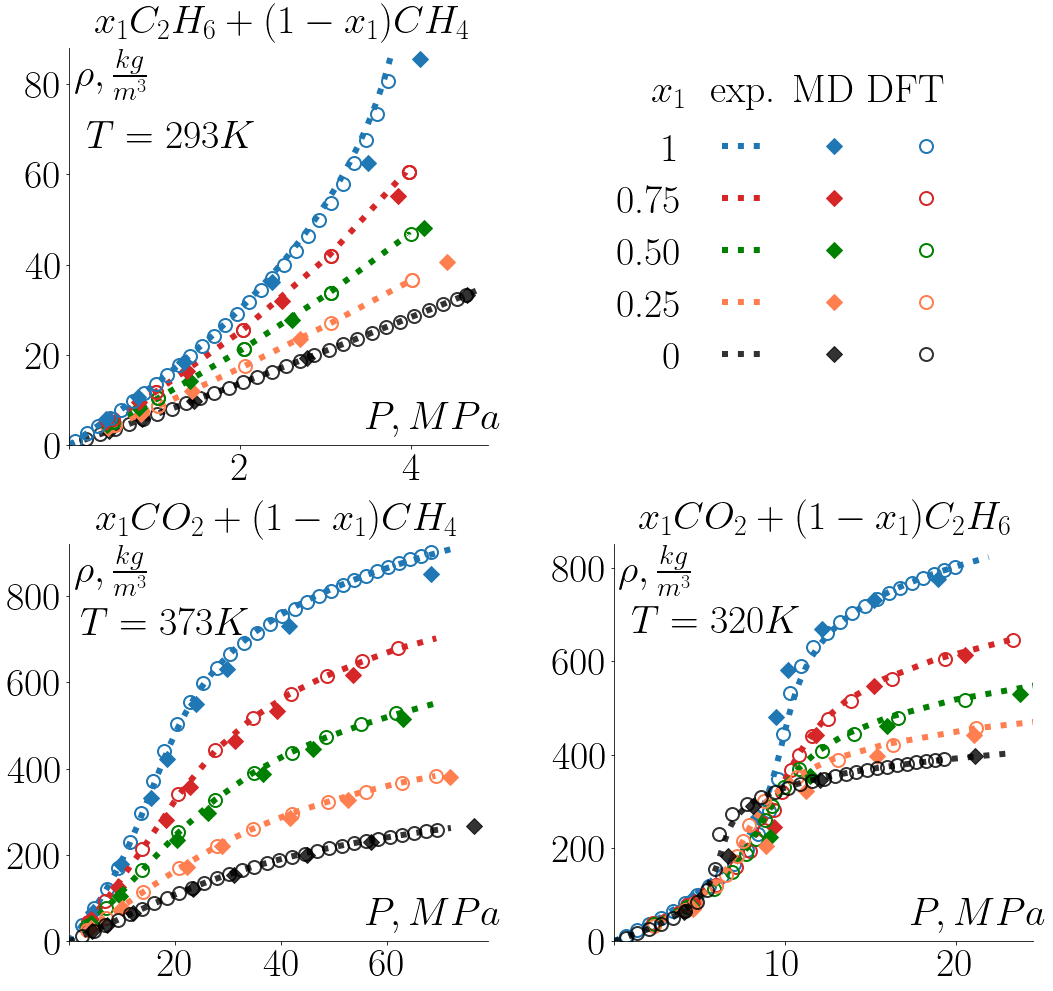}
\caption{Isotherms for binary mixtures in bulk. Validation of MD and DFT results on experimental data}
\label{fig:mix_bulk}
\end{figure}

Here again before linking MD and DFT to study fluid mixture under confinement, we validate theoretical and simulation models independently in bulk. We calculated isotherms for three binary mixtures (figure \ref{fig:mix_bulk}): ethane + methane at 293 K, carbon dioxide + methane at 373 K and ethane + carbon dioxide at 320 K. For the mixtures we considered the temperatures, pressure ranges and component concentrations for which experimental data are available in the literature \cite{humberg2018measurement, dewitt1966viscosities, diller1988measurements}. For pure fluids data from NIST \cite{nist} was used for comparison. Both theory and simulation reproduce the experimental dependencies well.

Finally, we present MD and DFT results on filling the pore with a fluid mixture for slit nanopores of different widths.

We considered three binary mixtures in 1, 3 and 5 (nm) wide pores under fixed conditions in bulk:
\begin{enumerate}
    \item 
    ethane + methane at 293~K, bulk pressure 2~MPa, bulk mole fraction of methane 75.12~\%;
    \item
    carbon dioxide + methane at 373~K, bulk pressure 10~MPa, bulk mole fraction of methane 75.5~\%;
    \item
    ethane + carbon dioxide at 320~K, bulk pressure 10~MPa, bulk mole fraction of ethane 74.8~\%.
\end{enumerate}

\begin{figure}[h!]
\centering
\includegraphics[width=\linewidth]{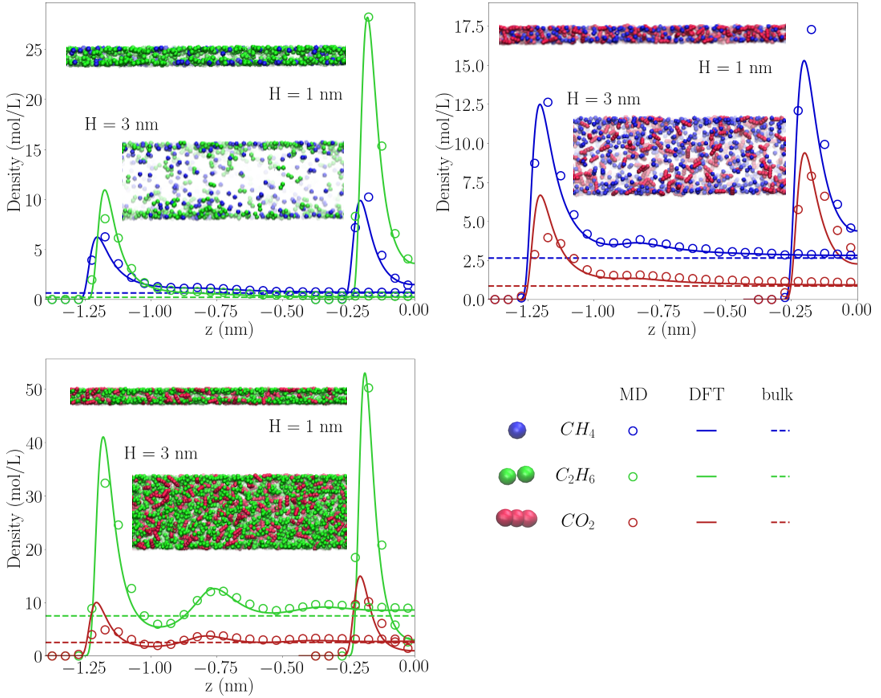}
\caption{MD and DFT density profiles for each component of binary mixtures}
\label{fig:mix_pore}
\end{figure}

Comparison of density profiles of each component obtained by MD and DFT is shown in figure \ref{fig:mix_pore} (for 1 and 3 nm pores). Profiles in 5 nm pores are identical to profiles in 3 nm pores except of the width of bulk-like region in the center. Only the half of the pore (from the wall to the pore center) is presented in the plots because of the symmetry of all profiles. 

For all mixtures considered the figures show good qualitative agreement. The difference between MD and DFT density profiles is similar to that for single-component fluids and is expected because of the different molecular models used by theory and simulation. The largest discrepancy is observed for carbon dioxide profiles which can be explained by non-isotropic molecules' orientation near the wall surface as described in section \ref{pure_pore}.

Mean percentage deviations between density profiles of each fluid component are shown in table \ref{tab:rmse_mix}. Here we also present the selectivity of the preferentially adsorbed component: $$S_{12} = \frac{x_{1}^{pore}/x_{2}^{pore}}{x_{1}^{bulk}/x_{2}^{bulk}}.$$

\begin{table}[h!]
    \centering
    \caption{NRMSD between MD and DFT density profiles for each component of binary mixtures and the adsorption selectivity of the first component}
    \label{tab:rmse_mix}
    \begin{tabular}{c|c|c|c|c}
    \multirow{2}{*}{Mixture} & \multirow{2}{*}{H, nm} & \multicolumn{2}{c|}{NRMSD, \%} &  \multirow{2}{*}{$S_{12}$} \\
    & & Comp.1 & Comp.2 & \\
    \hline
    \multirow{3}{*}{$C_2H_6 + CH_4$ } & 1 & 7.9 & 9.1 & 7.40 \\
    & 3 & 6.2 & 4.5 & 2.72 \\
    & 5 & 5.3 & 3.5 & 2.23 \\
    \hline
    \multirow{3}{*}{$CO_2 + CH_4$ } & 1 & 10.5 & 8.6 & 1.74 \\
    & 3 & 10.6 & 4.1 & 1.23 \\
    & 5 & 9.0 & 3.1 & 1.15 \\
    \hline
    \multirow{3}{*}{$C_2H_6 + CO_2$ } & 1 & 7.0 & 8.6 & 1.07 \\
    & 3 & 5.2 & 13.1 & 1.12 \\
    & 5 & 4.0 & 10.4 & 1.08 \\
\end{tabular}
\end{table}

For ethane/methane mixture we observe preferential adsorption of ethane over methane as a result of stronger interaction with carbon wall. Carbon dioxide is also selectively adsorbed over methane but its selectivity is lower under studied conditions. As we can see, in both cases (ethane/methane and carbon dioxide/methane) adsorption selectivity is higher in smaller pores which is consistent with the results reported in the literature \cite{kurniawan2006simulation, pitakbunkate2017phase}.
For the third mixture the selectivity of ethane over carbon dioxide does not vary significantly with pore width and is slightly higher than 1 in all cases considered.

\section{Conclusion}
This work provides the transition between theoretical (DFT) and simulation (MD) methods to study fluid behavior in nanopores in equilibrium with bulk. We propose performing MD simulation of the confined system, which is created based on the results of DFT calculation of the fluid composition and density distribution inside the pore under given bulk conditions. Such combination is computationally efficient and applicable to describe a wide range of phenomena (both equilibrium and dynamic) in nanoporous media.

We considered three single-component fluids: methane, ethane and carbon dioxide and three their binary mixtures in slit-like nanopores. Filling of pores of width 1, 3 and 5 (nm) at typical reservoir temperatures (273--373~K) and pressures (1.5--10~MPa) was studied. To validate the proposed approach we compared the equilibrium density profiles obtained by DFT and MD.

From the visual comparison of profiles 
we observe that both theory and simulation reproduce well the layering  structure of all studied fluids inside the pore. We also notice that both methods predict capillary condensation at the same bulk conditions.

For quantitative comparison mean percentage deviation between profiles of each component is calculated. The largest quantitative discrepancy is observed for carbon dioxide. The analysis of order parameter inside the pore in MD simulation shows that carbon dioxide molecules in the adsorption layer tend to orient parallel to the surface. This fact is not taken into account by the spherically symmetrical DFT model which explains the reason of the discrepancy.

In general, for nonpolar hydrocarbons (methane, ethane and their binary mixture) the agreement between the two methods used is better than for polar carbon dioxide (either pure or mixed with hydrocarbons). It shows that the electrostatic interactions in the confined fluid are important and should be considered in theory for more exact correspondence.

Presented results approve the applicability of the proposed method provided preliminary selection of accurate molecular models for both theory and simulation.  

In future the proposed approach can be used to investigate transport properties of confined fluids and phase transitions in nanopores. Another future direction is to expand this method to consider more realistic solid models and more complex fluids for conducting precise computational experiments.

\appendix

\setcounter{figure}{0} 
\section{Details of intermolecular interaction parameters}
\label{param_app}

To calculate intermolecular parameters for methane + ethane, we use Halgren HHG mixing rule (H--HHG):
\begin{equation}\label{eq:Hal-HHG}
    \sigma_{ij} = \frac{\sigma_{ii}^3 + \sigma_{jj}^3}{\sigma_{ii}^2 + \sigma_{jj}^2},
    \quad
    \epsilon_{ij} = \frac{4\epsilon_{ii}\epsilon_{jj}}{\epsilon_{ii}^{1/2} + \epsilon_{jj}^{1/2}}
\end{equation}

Adaptive Lorentz\,--\,Berthelot (ALB) rule is used to determine intermolecular paramreters for methane + carbon dioxide mixture:
\begin{equation}\label{eq:Ad-Lor-Bert}
    \sigma_{ij} = \frac{\sigma_{ii} + \sigma_{jj}}{2} \left(1-k_{ij} \right),
    \quad
    \epsilon_{ij} = \sqrt{\epsilon_{ii}\epsilon_{jj}} \left(1-l_{ij} \right)
\end{equation}
with coefficients $k_{ij}$ and $l_{ij}$ that are selected by fitting experimental data. For methane + carbon dioxide $k_{ij} = 0$,  $l_{ij} = 0.070913$.

To obtain ethane + carbon dioxide mixture parameters, we also used Adaptive Lorentz\,--\,Berthelot rule (ALB), but with $k_{ij} = 0$ and $l_{ij} = 0.098105$.

The choice of mixing rules is based on the results of \cite{nesterova2021adaptive} article.

\section*{Acknowledgements}
This research did not receive any specific grant from funding agencies in the public, commercial, or not-for-profit sectors.

 \bibliographystyle{elsarticle-num} 
 \bibliography{cas-refs}





\end{document}